\documentclass[twocolumn,tighten,times,twocolappendix,numberedappendix]{aastex631}

\usepackage{amsmath}
\usepackage{color}
\usepackage{url}
\usepackage{stmaryrd}

\usepackage{etoolbox}

\usepackage{graphicx}
\usepackage[flushleft]{threeparttable}
\usepackage{placeins}
\usepackage{comment}

\usepackage[encapsulated]{CJK}
\usepackage{ucs}
\usepackage[utf8x]{inputenc}
\newcommand{\cntext}[1]{\begin{CJK}{UTF8}{gbsn}#1\end{CJK}}
\newcommand{\japtext}[1]{\begin{CJK}{UTF8}{min}#1\end{CJK}}

\hypersetup{breaklinks=true,colorlinks=true,linkcolor=blue,citecolor=blue,filecolor=blue,urlcolor=blue}
\makeatletter
\patchcmd{\NAT@citex}
  {\@citea\NAT@hyper@{%
     \NAT@nmfmt{\NAT@nm}%
     \hyper@natlinkbreak{\NAT@aysep\NAT@spacechar}{\@citeb\@extra@b@citeb}%
     \NAT@date}}
  {\@citea\NAT@nmfmt{\NAT@nm}%
   \NAT@aysep\NAT@spacechar\NAT@hyper@{\NAT@date}}{}{}

\patchcmd{\NAT@citex}
  {\@citea\NAT@hyper@{%
     \NAT@nmfmt{\NAT@nm}%
     \hyper@natlinkbreak{\NAT@spacechar\NAT@@open\if*#1*\else#1\NAT@spacechar\fi}%
       {\@citeb\@extra@b@citeb}%
     \NAT@date}}
  {\@citea\NAT@nmfmt{\NAT@nm}%
   \NAT@spacechar\NAT@@open\if*#1*\else#1\NAT@spacechar\fi\NAT@hyper@{\NAT@date}}
  {}{}
  
\usepackage{array}
\newcolumntype{C}[1]{>{\centering\let\newline\\\arraybackslash\hspace{0pt}}m{#1}}
\usepackage{xspace}

\def\aj{AJ}

\def\apj{ApJ}
\def\apjl{ApJ}
\def\apjs{ApJS}
\def\ao{Appl.~Opt.}

\def\aap{A\&A}
\def\aapr{A\&A~Rev.}
\def\aaps{A\&AS}

\def\mnras{MNRAS}

\def\pasp{PASP}

\def\qjras{QJRAS}

\def\iaucirc{IAU~Circ.}

\def\arcdeg{\hbox{$^\circ$}}

\definecolor{burgundy}{rgb}{0.5, 0.0, 0.13}

\usepackage{xspace}

\newcommand{\orcidicon}{\includegraphics[width=0.26cm]{orcid-ID.eps}}

\newcommand{\orcidauthor}[1]{\href{https://orcid.org/#1}{\orcidicon}}

\shorttitle{ALMA observations of V\,605~Aql}
\shortauthors{Tafoya et al.}

\makeatletter
\patchcmd{\frontmatter@RRAP@format}{(}{}{}{}
\patchcmd{\frontmatter@RRAP@format}{)}{}{}{}
\renewcommand\Dated@name{}
\makeatother

\begin{document}

\title{First images of the molecular gas around a born-again star revealed by ALMA}

\correspondingauthor{Daniel\,Tafoya}
\email{daniel.tafoya@chalmers.se}

\author[0000-0002-2149-2660]{Daniel\,Tafoya\japtext{(多穂谷)}}
\affil{Department of Space, Earth and Environment, Chalmers University of 
Technology, Onsala Space Observatory, SE-439 92 Onsala, Sweden}

\author[0000-0002-5406-0813]{Jes\'{u}s\,A.\,Toal\'{a}\cntext{(杜宇君)}}
\affil{Instituto de Radioastronom\'ia y Astrof\'isica,  UNAM 
Campus Morelia, Apartado postal 3-72, 58090, Morelia, Michoacán, Mexico}

\author[0000-0002-2843-2476]{Ramlal\,Unnikrishnan}
\affil{Department of Space, Earth and Environment, Chalmers University of 
Technology, Onsala Space Observatory, SE-439 92 Onsala, Sweden}

\author[0000-0002-2700-9916]{Wouter~H.~T.\,Vlemmings}
\affil{Department of Space, Earth and Environment, Chalmers University of 
Technology, Onsala Space Observatory, SE-439 92 Onsala, Sweden}

\author[0000-0002-7759-106X]{Mart\'{i}n~A.\,Guerrero}
\affil{Instituto de Astrof\'{i}sica de Andaluc\'{i}a, IAA-CSIC, Glorieta de la Astronom\'{i}a S/N, E-18008 Granada, Spain}

\author[0000-0003-2379-0474]{Stefan\,Kimeswenger}
\affiliation{Institut f{\"u}r Astro- und Teilchenphysik, Universit{\"a}t Innsbruck, Technikerstr. 25\/8, 6020 Innsbruck, Austria}
\affiliation{Instituto de Astronom{\'i}a, Universidad Cat\'olica del Norte, Av. Angamos 0610, Antofagasta, Chile}

\author[0000-0001-7490-0739]{Peter~A.~M.\,van Hoof}
\affil{Royal Observatory of Belgium, Ringlaan 3, B-1180 Brussels, Belgium}

\author[0000-0003-2343-7937]{Luis~A.\,Zapata}
\affil{Instituto de Radioastronom\'ia y Astrof\'isica,  UNAM 
Campus Morelia, Apartado postal 3-72, 58090, Morelia, Michoacán, Mexico}

\author[0000-0002-4033-2881]{Sandra~P.\,Trevi\~{n}o-Morales}
\affil{Department of Space, Earth and Environment, Chalmers University of Technology, Onsala Space Observatory, SE-439 92 Onsala, Sweden}

\author[0000-0002-0616-8336]{Janis~B.\,Rodr\'{i}guez-Gonz\'{a}lez}
\affil{Instituto de Radioastronom\'ia y Astrof\'isica,  UNAM 
Campus Morelia, Apartado postal 3-72, 58090, Morelia, Michoacán, Mexico}

\date[ ]{Submitted to ApJL}

\begin{abstract}
Born-again stars allow probing stellar evolution in human timescales and provide the most promising path for the formation of hydrogen-deficient post-asymptotic giant branch objects, but their cold and molecular components remain poorly explored. Here we present ALMA observations of V\,605~Aql that unveil for the first time the spatio-kinematic distribution of the molecular material associated to a born-again star. Both the continuum and molecular line emission exhibit a clumpy ring-like structure with a total extent of $\approx$1$^{\prime\prime}$ in diameter. The bulk of the molecular emission is interpreted as being produced in a radially-expanding disk-like structure with an expansion velocity v$_{\rm exp}$$\sim$90~km~s$^{-1}$ and an inclination $i$$\approx$60$^{\circ}$ with respect to the line-of-sight. The observations also reveal a compact high-velocity component, v$_{\rm exp}$$\sim$280~km~s$^{-1}$, that is aligned perpendicularly to the expanding disk. This component is interpreted as a bipolar outflow with a kinematical age $\tau$$\lesssim$20~yr, which could either be material that is currently being ejected from V\,605~Aql, or it is being dragged from the inner parts of the disk by a stellar wind. The dust mass of the disk is in the range $M_{\rm dust}$$\sim$0.2--8$\times$10$^{-3}$~M$_{\odot}$, depending on the dust absorption coefficient. The mass of the CO is $M_{\rm CO}$$\approx$1.1$\times10^{-5}$~$M_{\odot}$, which is more than three orders of magnitude larger than the mass of the other detected molecules. We estimate a $^{12}$C/$^{13}$C ratio of 5.6$\pm$0.6, which is consistent with the single stellar evolution scenario in which the star experienced a very late thermal pulse instead of a nova-like event as previously suggested.  
\end{abstract}


\keywords{\href{https://astrothesaurus.org/uat/847}{Interstellar medium (847)};
\href{https://astrothesaurus.org/uat/1249}{Planetary nebulae (1249)};
\href{http://astrothesaurus.org/uat/1607}{Stellar jets (1607)};
\href{http://astrothesaurus.org/uat/1636}{Stellar winds (1636)};
\href{http://astrothesaurus.org/uat/2050}{Low mass stars(2050)}
\vspace{4pt}
\newline
}


\section{Introduction}
\label{wc:sec:introduction}

Born-again stars are post-asymptotic giant branch (post-AGB) objects that are thought to have experienced a final helium shell flash. This event occurs when the burning ashes of the outermost hydrogen layer produce a helium layer that reaches the critical mass to ignite thermonuclear reactions and make carbon and oxygen \citep{Iben1983}. This is considered the most promising mechanism to explain the existence of hydrogen-deficient post-AGB objects. During this process, material from the star is violently ejected into the surroundings. After temporally re-visiting the locus of AGB stars in the Herzsprung–Russell (H–R) diagram, the born-again star reheats and becomes hot enough to ionize the ejected material, forming a hydrogen-deficient, helium- and carbon-rich planetary nebula \citep[PN; sometimes referred to as ``born-again PN''; e.g.,][]{Guerrero2012} within the old hydrogen-rich PN.  If the final helium shell flash occurs when the star is already on the white dwarf cooling track, which is known as a very late thermal pulse (VLTP), and the old PN is still visible, the born-again PN will show up in optical images as a bright compact PN nested inside a faint extended PN \citep[][and references therein]{Jacoby1979,Jacoby1983,Pollacco1999,Guerrero1996,Gvaramadze2020}. Even though it is thought that 10\% to $\sim$25\% of stars that evolve off the AGB phase experience a final flash, due to its brevity, a VLTP has been directly observed only in two objects: V\,605~Aql and V\,4334~Sgr (the latter is also known as Sakurai's Object).

Shortly after the VLTP, the effective temperature of the born-again star drops $T_{\rm eff}$$\lesssim$5000~K as predicted by stellar evolution models \citep[e.g.,][]{Herwig1999,MB2006}. Thus, optimal conditions for the formation of dust and molecules are expected as a consequence of the born-again phenomenon. Given the high content of carbon of the ejecta, it shortly condenses into carbon-rich dust grains  \citep[see][]{Clayton1997,Evans2020}. Such swift formation of dust and molecules has not been addressed by chemical evolution models, but their properties have been extensively observationally studied at optical and IR wavelengths \citep[e.g.,][and references therein]{Borkowski1994,Cohen1977,Clayton2013,Koller2001,Toala2021}. In the case of Sakurai's Object, which experienced the VLTP $\sim$30~yr ago, several molecular species have been detected \citep{Evans2020}. \citet{Eyres1998} reported the presence of C$_{2}$, CN and $^{12}$CO features in near-IR spectra taken just one year after the born-again event, and \citet{Evans2006} reported the presence of hydrogenated molecules using {\it Spitzer} mid-IR spectroscopic observations \citep[see][and references therein]{Hinkle2020}. In the sub-millimeter regime, \citet{Tafoya2017} reported the detection of molecular emission associated with born-again stars. From their APEX observations, emission of the HCN($J$$=$4$\shortrightarrow$3) and H$^{13}$CN($J$$=$4$\shortrightarrow$3) lines was detected toward Sakurai's object, while CO($J$$=$3$\shortrightarrow$2) emission was detected toward V\,605~Aql. These observations revealed that the molecular gas is expanding at relatively high-velocities ($\gtrsim$80~km~s$^{-1}$), but it was not possible to associate the molecular gas to the different nebular components.

\begin{figure*}
	\begin{center}
		\includegraphics[width=0.95\linewidth]{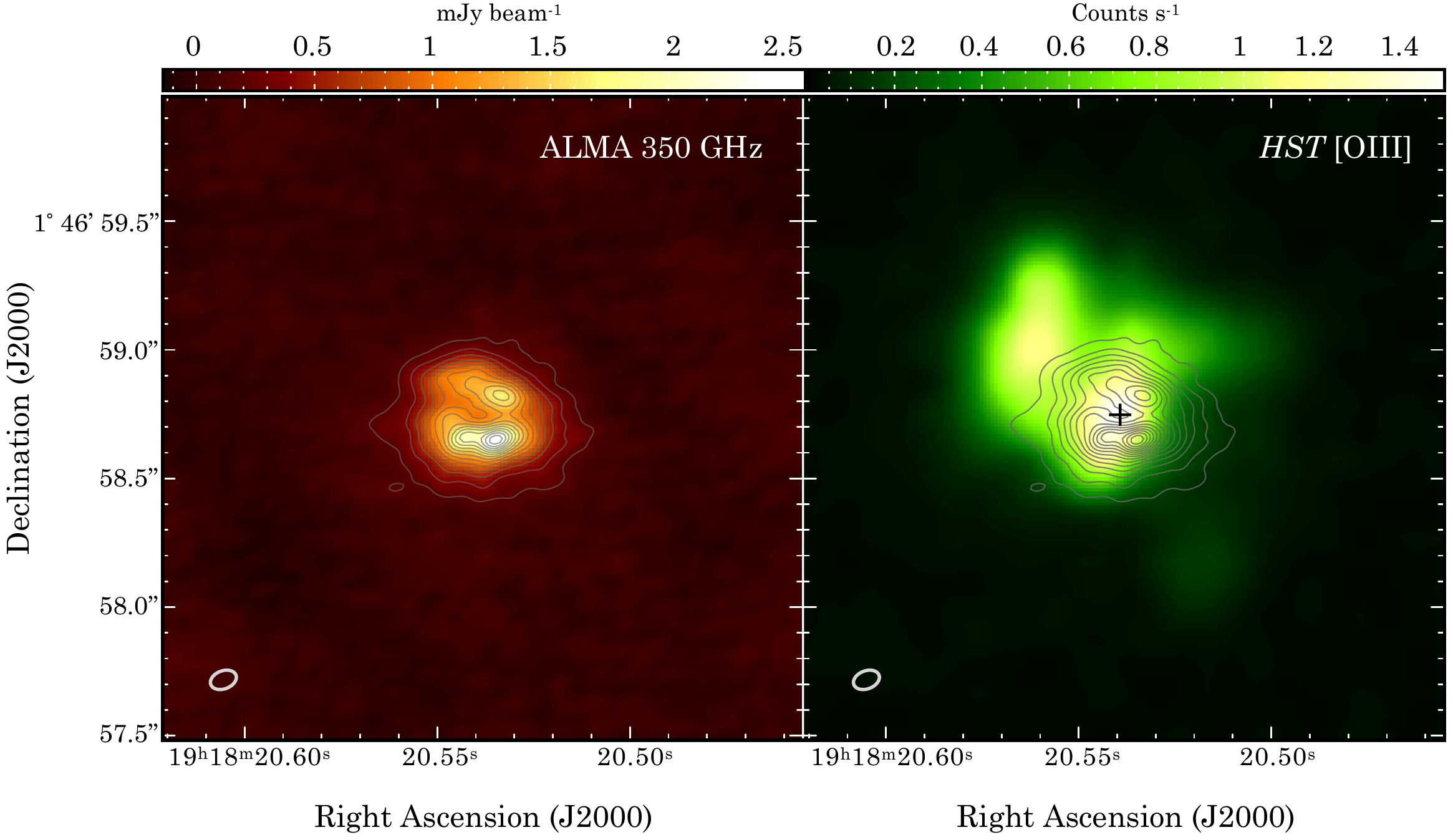}
		\caption{Comparison between the ALMA continuum image and the optical {\it HST} [O\,{\sc iii}] image of V\,605~Aql obtained on 2009 (PI: G.\,Clayton; Program ID: 11985). {\bf Left:} Image of the ALMA continuum emission at 350~GHz. The contours are drawn from 5-$\sigma$ on steps of 5-$\sigma$ (where $\sigma$ is the rms noise level of 35~$\mu$Jy~beam$^{-1}$). {\bf Right:} The color map is the optical {\it HST} [O\,{\sc iii}] image taken on 2009 March, and the contours are the same as in the left panel. A black cross marks the approximate position of the central star as found by \citet{Clayton2013}: (J2000) R.A.$=$19$^{\rm h}$18$^{\rm m}$20$\rlap{.}^{\rm s}$538, Dec.$=$+1$^{\circ}$46$^{\prime}$58$\rlap{.}^{\prime\prime}$741. In both panels, the synthesized beam of the ALMA observations is shown in the bottom-left corner, and its parameters are: $\theta_{\rm beam}$$=$0$\rlap{.}^{\prime\prime}$107$\times$0$\rlap{.}^{\prime\prime}$070, P.A.$=$$-$69.1$^{\circ}$.}
		\label{fig:continuum}
	\end{center}
\end{figure*}

Optical observations have revealed that during the born-again event highly-processed (hydrogen-deficient, helium- and carbon-rich) material is expelled at relatively high speeds \citep[$\sim$40--300~km~s$^{-1}$;][]{Meaburn1996,Meaburn1998,vanHoof2018,Toala2021_HuBi1}. From images of the more evolved born-again PNe, A30 and A78, it has been seen that the spatial distribution of the ejecta consists of a toroidal structure, or a disrupted disk, and a pair of knots expanding perpendicularly to the disk \citep[][]{Fang2014}. A similar morphology has been observed in the youngest born-again star Sakurai's Object \citep{Hinkle2014,Hinkle2020}. {\it HST} images show that V\,605~Aql, which underwent the VLTP event in 1919, about a hundred years ago, also exhibits a bipolar morphology with hints of a toroidal component \citep[][]{Hinkle2008,Clayton2013}. Its central star has experienced dramatic changes, evolving from a $T_\mathrm{eff}\approx$5000~K in 1921 to become a hot $T_\mathrm{eff}>$90,000~K carbon-rich [Wolf-Rayet] star by 2001 \citep{Clayton2006}.

The study of the spatio-kinematical distribution of the ejected material is important because it can help to reveal how the VLTP took place. In this regard, observations of molecular emission have proven to be a powerful tool to study the morphology and kinematics, as well as 
the physical conditions of circumstellar material in a wide variety of astrophysical contexts. However, due to their relatively small size, to date there are no resolved images of the distribution of the molecular material around the two youngest born-again stars. For the more evolved born-again stars, no molecular emission has been detected and only studies of the dust have been conducted. In this paper we present the first images of the molecular emission and sub-millimeter continuum emission around the born-again star V\,605~Aql. The distance to V\,605~Aql has been estimated around through various methods resulting in a whole set of distances between 3.1 and 5.7~kpc \citep[see e.g.,][and references therein]{Maciel1984,Lechner2004,Clayton2013}. In this paper we adopt a canonical mean value of 4~kpc.

\begin{figure*}
	\begin{center}
		\includegraphics[width=0.7\linewidth]{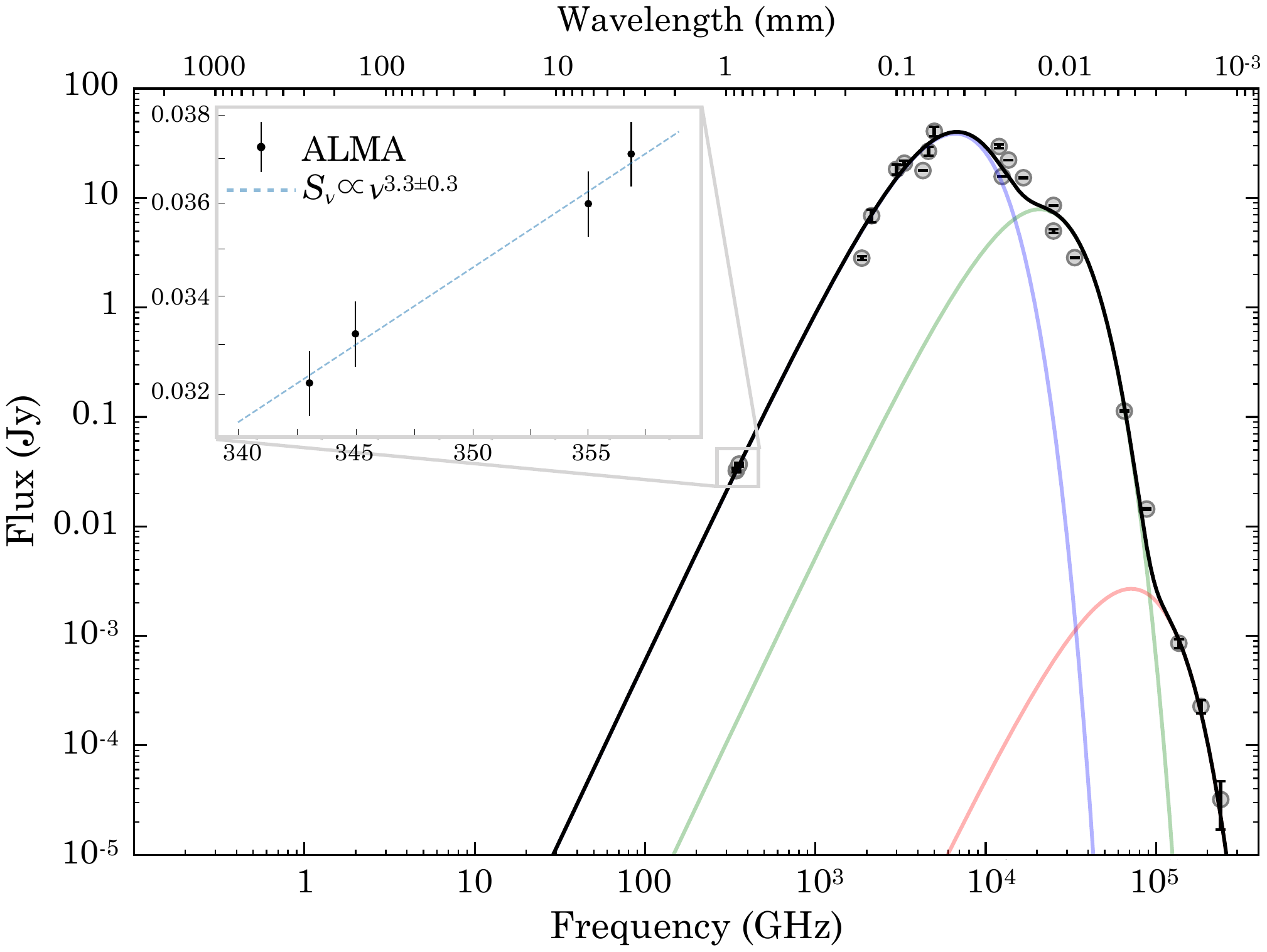}
	    \caption{Spectral energy distribution (SED) of V\,605~Aql. All data points, except those from the ALMA observations, were taken from Table~2 of \citet[][]{Clayton2013}. The bars of the ALMA data points, shown in the inset, represent flux density error of 5\%. A power-law fit to the ALMA data gives $S_{\nu}$$\propto$$\nu^{3.3\pm0.3}$ and it is shown with a dashed blue line in the inset. The units of the inset are the same as the main panel. The red, green and blue solid lines correspond to modified black bodies at temperatures 810~K, 235~K and 75~K, respectively. The emissivity spectral index of the modified black bodies is $\beta$$=$1.3.}
		\label{fig:SED}
	\end{center}
\end{figure*}

\begin{figure*}
	\begin{center}
		\includegraphics[width=0.8\linewidth]{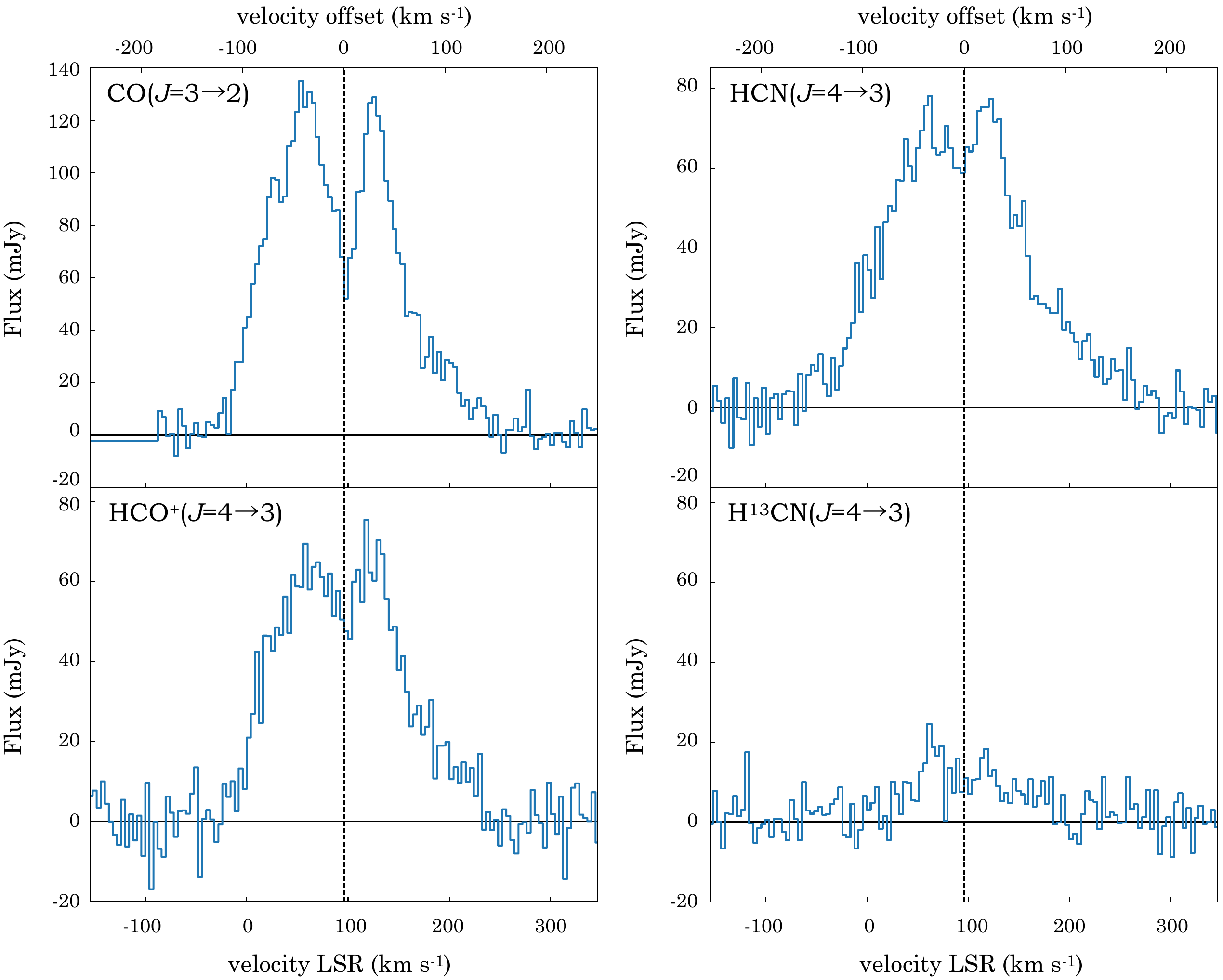}
		\caption{Spectra of the molecular emission detected toward V\,605~Aql from the ALMA observations integrated over a circular region centered at the position of the central star and with a diameter of 1$\rlap{.}^{\prime\prime}$2. The width of the channels is 4~km~s$^{-1}$. The dashed vertical line corresponds to the systemic velocity relative to the local standard of rest of V\,605~Aql, v$_{\rm{sys,LSR}}$$=$96~km~s$^{-1}$ \citep{Tafoya2017}.}
		\label{fig:spectra}
	\end{center}
\end{figure*}

\section{Observations}\label{sec:observations}

We used the Atacama Large Millimeter/submillimeter Array (ALMA) to observe with high angular resolution 
(0$\rlap{.}^{\prime\prime}$07$\times$0$\rlap{.}^{\prime\prime}$1) the continuum and molecular emission from the material around the born-again star V\,605~Aql (project 2019.1.01408.S; PI: D.\,Tafoya). The observations were carried out as three separate sessions on 2021 July 4, 10 and 11, using 42, 46 and 45 antennas, respectively, of the ALMA 12~m array with Band 7 receivers ($\sim$350~GHz). The minimum and maximum baseline lengths were 15.3 m and 3.6 km, which provide a nominal angular resolution and maximum recoverable scale of 0$\rlap{.}^{\prime\prime}$09 and 1$\rlap{.}^{\prime\prime}$2, respectively. The field of view is $\sim$18$^{\prime\prime}$. The total time on the science target was 1h and 56 min. The average precipitable water vapor level during the observations was around 0.6 mm. 

The data were calibrated using the ALMA pipeline (version 2020.1.0.40; CASA 6.1.1.15) with J1924$-$2914 as the amplitude and bandpass calibrator (flux density$=$2.4~Jy at 355~GHz and spectral index $\alpha$$=$$-$0.586). Phase calibration was done using J1851+0035 (flux density$=$279~mJy at 355~GHz). Atmospheric variations during the observations were corrected by using water vapor radiometer data. The calibrated data set contained 4$\times$1.85 GHz spectral windows with 2000 channels each.

Images were created with CASA~5.6.2-3 \citep{McMullin2007} using a Briggs weighting scheme with the robust parameter set to 0.5. The continuum emission was subtracted from the data cubes and channel maps with a spectral resolution of 4~km~s$^{-1}$ were created. The typical root-mean-square (rms) noise level in each line-free channel is $\sim$0.5~mJy~beam$^{-1}$. A continuum image was created applying multi-frequency synthesis \citep{Rau2011} on line-free channels from the four spectral windows (spws), giving a total effective bandwidth of 4.73~GHz. The central frequencies of the spws used for creating the continuum are 343.031~GHz, 344.989~GHz, 355.031~GHz and 356.911~GHz, respectively. The rms level in the continuum image is $\sim$35~$\mu$Jy~beam$^{-1}$.

\begin{figure*}
	\begin{center}
		\includegraphics[width=0.9\linewidth]{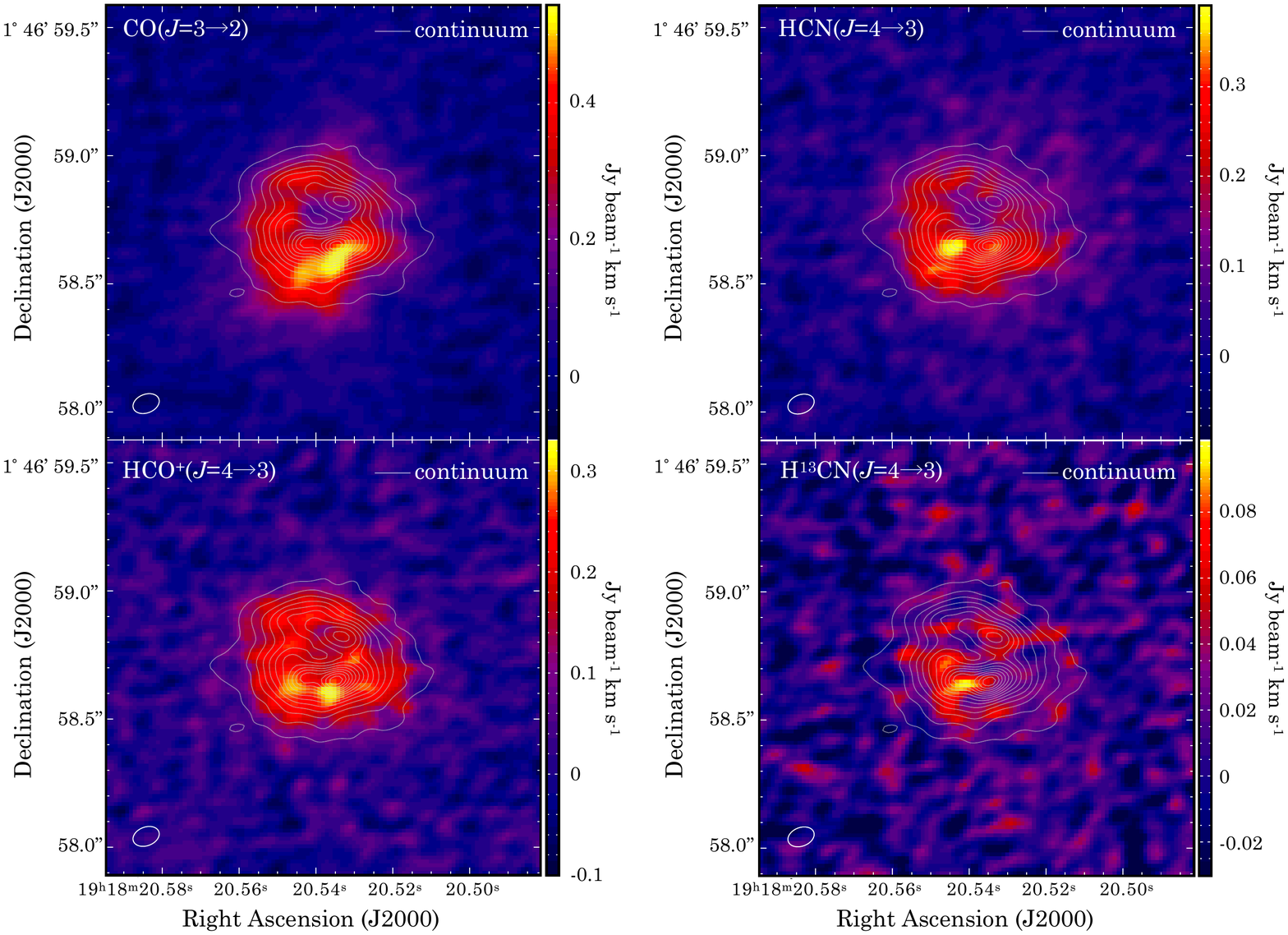}
		\caption{Velocity-integrated emission, zeroth moment, of the molecular emission detected toward V\,605~Aql from the ALMA observations. The images were obtained by integrating all channels with line emission above a 3-$\sigma$ level, with $\sigma$=0.5~mJy~beam$^{-1}$. The contours and parameters of the synthesized beam are the same as in Fig.~\ref{fig:continuum}. The rms noise level of the zeroth moment images is $\sim$21~mJy~beam$^{-1}$~km~s$^{-1}$.}
		\label{fig:moment0}
	\end{center}
\end{figure*}

\section{Results and Discussion}
\label{sec:results}

\subsection{Sub-millimeter continuum and molecular line emission} \label{sec:Sub-millimeter_continuum_and_molecular_line_emission}

From the ALMA observations we obtained the first images of the sub-millimeter continuum and molecular emission of the material ejected by the born-again star V\,605~Aql. The continuum emission appears as a broken and clumpy ring-like structure with an average diameter of $\approx$0$\rlap{.}^{\prime\prime}$3, although there is fainter emission extending over a region of $\approx$0$\rlap{.}^{\prime\prime}$7 in diameter. The ring-like structure is slightly elongated in the East-West direction (see Fig.~\ref{fig:continuum}). In the right panel of Figure~\ref{fig:continuum} the ALMA continuum image is compared to the [O\,{\sc iii}] emission from {\it HST} observations. The {\it HST} image was shifted 0$\rlap{.}^{\prime\prime}$3 in Declination to align the geometric center of the continuum ring, (J2000) R.A.$=$19$^{\rm h}$18$^{\rm m}$20$\rlap{.}^{\rm s}$538, Dec.$=$+1$^{\circ}$46$^{\prime}$58$\rlap{.}^{\prime\prime}$741, with the [O\,{\sc iii}] emission peak, which is approximately the location of the central star found by \citet{Clayton2013}. The continuum emission is less extended than the [O\,{\sc iii}] emission and below we propose that it is associated to the dark band seen at optical wavelengths.

\begin{figure*}
\begin{center}
\includegraphics[width=0.85\linewidth]{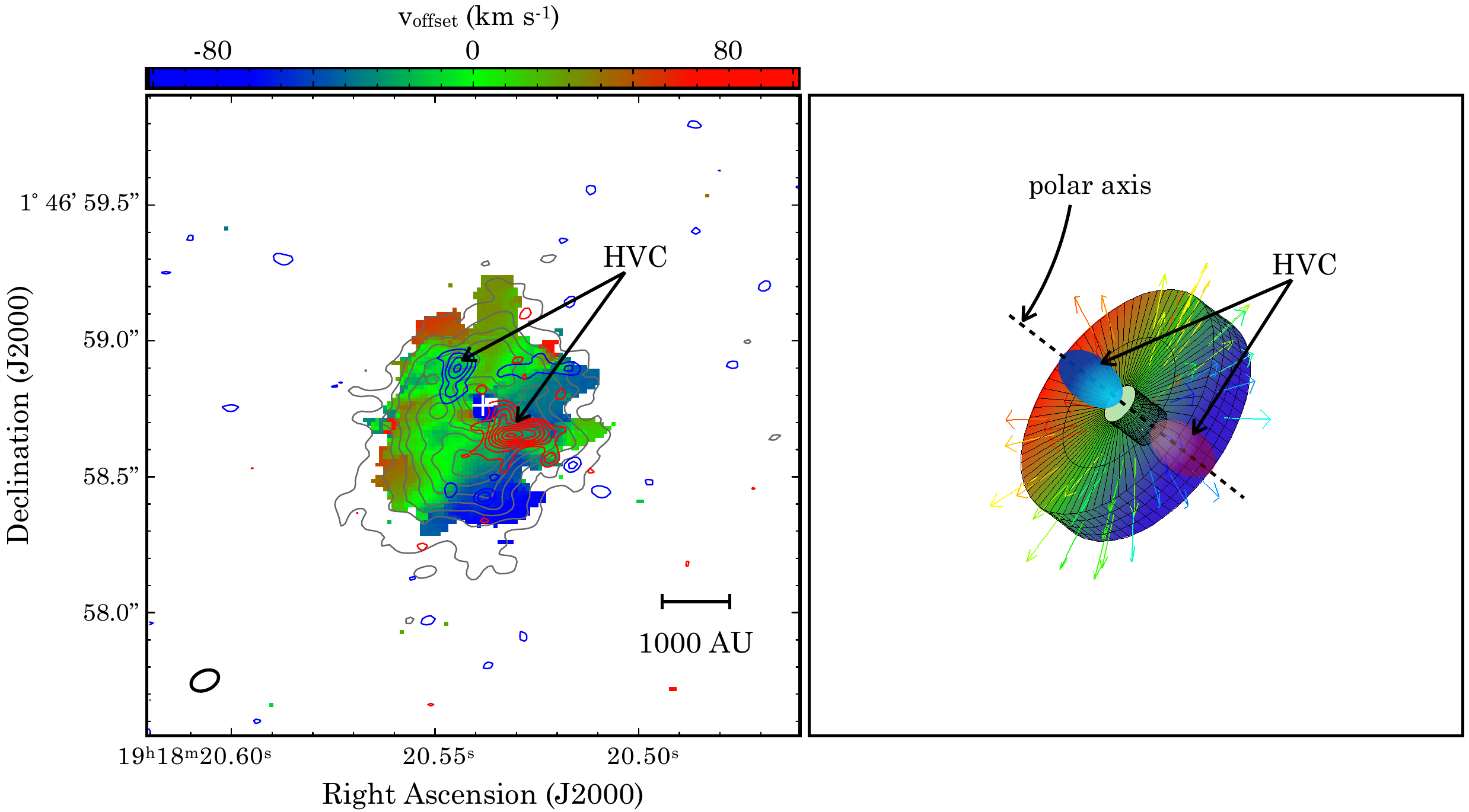}
\caption{Spatio-kinematical distribution of the CO($J$$=$3$\shortrightarrow$2) emission in V\,605~Aql. {\bf Left}: The velocity field (first moment) of the CO($J$$=$3$\shortrightarrow$2) emission in the velocity range $-$100$\lesssim$v$_{\rm offset}$(km~s$^{-1}$)$\lesssim$+100 is shown as a color map. The pixels with emission below 3 times the rms noise level of 0.5~mJy~beam$^{-1}$ were masked. The gray contours show the velocity-integrated emission (zeroth moment) emission of the CO($J$$=$3$\shortrightarrow$2) in the velocity range $-$100$\lesssim$v$_{\rm offset}$(km~s$^{-1}$)$\lesssim$+100. The gray contours are drawn from 3-$\sigma$ on steps of 3-$\sigma$ (where $\sigma$$=$21~mJy~beam$^{-1}$~km~s$^{-1}$ is the rms noise level of the zeroth moment image). The blue and red contours (HVC) show the zeroth moment emission of the CO($J$$=$3$\shortrightarrow$2) in the velocity range $-$120$\gtrsim$v$_{\rm offset}$(km~s$^{-1}$)$\gtrsim$-100 and +100$\lesssim$v$_{\rm offset}$(km~s$^{-1}$)$\lesssim$+140, respectively. The blue and red contours are drawn from 3-$\sigma$ on steps of 3-$\sigma$ (where $\sigma$$=$7~mJy~beam$^{-1}$~km~s$^{-1}$ is the rms noise level of the zeroth moment image). The white cross has the same coordinates as the black cross of Fig.\ref{fig:continuum}. The horizontal bar indicating the linear scale of the image assumes a distance of 4~kpc to the source. The parameters of the synthesized beam, shown in the bottom-left corner, are the same as in Fig.~\ref{fig:continuum}. {\bf Right}: Spatio-kinematical model of the molecular emission of V\,605~Aql produced with the software SHAPE.}
\label{fig:moment1_model}
\end{center}
\end{figure*}

Apart from the continuum image that was created by combining the emission in the four spws, continuum images of each individual spw were also created. As mentioned in section~\ref{sec:observations}, the central frequencies of the spws are 343.031~GHz, 344.989~GHz, 355.031~GHz and 356.911~GHz, and their corresponding continuum emission fluxes are 32.2~mJy, 33.2~mJy, 36.0~mJy and 37.1~mJy. A power-law fit to the spectral energy distribution (SED) in the frequency range of the ALMA observations gives $S_{\nu}$$\propto$$\nu^{3.3\pm0.3}$ (see Fig.~\ref{fig:SED}). It is well known that at sub-mm wavelengths dust is the most important source of continuum emission, whose flux exhibits a power law dependence with frequency as $S_{\nu}$$\propto$$\nu^{2+\beta}$, where $\beta$ is the emissivity spectral index of the dust grains \citep{Hildebrand1983}. Thus, the power-law dependence of the ALMA continuum corresponds to dust emission with an emissivity spectral index $\beta$$=$1.3$\pm$0.3. Figure \ref{fig:SED} shows the SED of V\,605~Aql including observations within the wavelength range from $\sim$1$\mu$m to $\sim$1mm. Similarly to \citet[][]{Clayton2013}, we did a three component fit to the data points, including our ALMA observations, but we used modified black body curves with an emissivity spectral index $\beta$$=$1.3. The 75~K component fits very well the continuum emission at $\sim$350~GHz observed with ALMA. The dust mass can be estimated assuming optically thin emission and using the following expression:
\begin{equation*}
M_{\rm dust}=\frac{S_{\nu}\,D^{2}}{\kappa_{\nu}\,B_{\nu}(T_{\rm dust})}, 
\end{equation*}
\noindent
where $S_{\nu}$ is the flux density of the continuum emission, $D$ is the distance to the source, $\kappa_{\nu}$ is the dust absorption coefficient and $B_{\nu}(T)$ is the Planck function. The composition of the dust grains in V\,605~Aql is considered to be amorphous carbon \citep[see e.g.,][and references therein]{Clayton2013}. The value for the dust absorption coefficient of amorphous carbon reported in the literature ranges from $\kappa$($\nu$$=$350\,GHz)$\sim$1.3~cm$^{2}$\,gm$^{-1}$ up to $\kappa$($\nu$$=$350\,GHz)$\sim$60~cm$^{2}$\,gm$^{-1}$ \citep[e.g.,][]{Draine1984,Martin1987,Ossenkopf1994,Zubko1996,Mennella1998,Suh2000}, with values at the low end being more common. Thus, for a dust temperature $T_{\rm dust}$$=$75~K and a distance to the source $D$$=$4~kpc, the resulting dust mass varies in the range $M_{\rm dust}$$\sim$0.2--8$\times$10$^{-3}$~M$_{\odot}$, which is in agreement with the values, re-scaled by the distance, obtained by \citet[][]{Clayton2013} and \citet[][]{Koller2001}. It is clear that the adoption of different values for the dust absorption coefficient and distance to the source lead to significant differences in the derived values of the dust mass. Further work on determining the detailed characteristics of the dust grains in the ejecta of V\,605~Aql is necessary to better constrain the dust mass.   

Emission from the CO($J$$=$3$\shortrightarrow$2), HCN($J$$=$4$\shortrightarrow$3), H$^{13}$CN($J$$=$4$\shortrightarrow$3) and HCO$^{+}$($J$$=$4$\shortrightarrow$3) lines was also detected from the ALMA observations. Figure~\ref{fig:spectra} shows the spectra of the emission integrated over a circular region centered at the position of the central star and with a diameter of 1$\rlap{.}^{\prime\prime}$2. The peak flux of the CO($J$$=$3$\shortrightarrow$2) line agrees well with the APEX observations presented by \citet{Tafoya2017}, indicating that there is no missing flux from the interferometric observations. The lines exhibit a double peak profile and the emission extends over a relatively broad velocity range. The HCN($J$$=$4$\shortrightarrow$3) line seems to extend over a broader velocity range than the rest of the lines. The dip at the center of the lines coincides with the systemic velocity, v$_{\rm{sys,LSR}}$$=$96~km~s$^{-1}$, derived from the APEX observations. The velocity integrated intensity of the red-shifted side of the lines is lower than that of the blue-shifted side, although the peak emission is comparable.   

Figure~\ref{fig:moment0} shows the velocity-integrated emission, zeroth moment, images of the emission for the different molecular species.  The zeroth moment images were obtained by integrating the emission of the corresponding spectral line for each pixel along the velocity-axis within the velocity range $-$v$_\mathrm{max}$$\lesssim$v$_\mathrm{offset}$$\lesssim$$+$v$_\mathrm{max}$, where v$_{\rm offset}$ is defined as v$_\mathrm{offset}$$=$v$_\mathrm{LSR}\footnote{LSR stands for local standard of rest, which is an inertial reference frame based on the average velocity of stars in the solar neighborhood.} -$v$_\mathrm{sys,LSR}$, and v$_{\rm max}$ is the maximum velocity offset of the line. As it can be seen from Figure~\ref{fig:spectra}, v$_{\rm max}$ differs from one molecular species to another, varying from $\sim$150~km~s$^{-1}$ to $\sim$200~km~s$^{-1}$, the latter corresponding to the HCN($J$$=$4$\shortrightarrow$3) line. 

The molecular emission also exhibits a clumpy ring-like structure and its spatial extent is somehow similar to that of the continuum emission, shown as white contours in the panels of Figure~\ref{fig:moment0}. Particularly, the CO\,($J$$=$3$\shortrightarrow$2) line is slightly more extended than the continuum emission and it is elongated in a different direction (see section \ref{sec:expanding_disk}). The molecular ring is brighter in the southern edge, where the continuum emission is brighter too. We calculated the mass of the gas for the different molecular species assuming optically thin emission, LTE conditions, and adopting a distance of 4~kpc. For these calculations we followed a formalism similar to the one presented by \citet{Mangum2015} and used values of the Einstein A-coefficients, statistical weights, and partition functions from the Cologne Database for Molecular Spectroscopy \citep{Muller2001,Muller2005}. For an excitation temperature of $T_{\rm ex}$$=$75~K, which is the temperature of the dust component that fits the ALMA continuum, the derived values of the molecular masses are $M_{\rm CO}$$=$1.05$\pm$0.01$\times$10$^{-5}$~$M_{\odot}$, $M_{\rm HCN}$$=$9.1$\pm$0.1$\times$10$^{-9}$~$M_{\odot}$, $M_{{\rm H}^{13}{\rm CN}}$$=$1.6$\pm$0.1$\times$10$^{-9}$~$M_{\odot}$ and $M_{{\rm HCO}^{+}}$$=$4.5$\pm$0.1$\times$10$^{-9}$~$M_{\odot}$. Thus, the main component of the molecular material is CO, perhaps even exceeding H$_{2}$ (although empirical determination of the H$_{2}$ content in born-again PNe is lacking).

\subsection{An expanding disk and a high-velocity outflow in V\,605~Aql} \label{sec:expanding_disk}

The velocity field of the molecular gas is obtained by calculating the intensity-weighted spectral moment distribution, first moment, of the emission. We created first moment images from the emission of all the detected molecules. Since the first moment of all the detected molecules exhibit similar overall spatio-kinematical distributions, we show in Figure~\ref{fig:moment1_model} the results for the CO($J$$=$3$\shortrightarrow$2) line, which is the dominant component of the molecular gas and the one with highest signal-to-noise ratio.

The first moment image was created including all the CO($J$$=$3$\shortrightarrow$2) emission in the velocity range $-$100$\lesssim$v$_{\rm offset}$(km~s$^{-1}$)$\lesssim$+100 and masking pixels with emission below 3 times the rms noise level of the channel maps. The velocity field exhibits a clear velocity gradient along the NE-SW direction from v$_{\rm offset}$$=$+80~km~s$^{-1}$ to v$_{\rm offset}$$=$$-$80~km~s$^{-1}$, resembling the typical velocity field of radially-expanding disks --and/or tori-- inclined with respect to the line-of-sight. On the other hand, the emission with velocity offset 80$\lesssim$$|$v$_{\rm offset}$(km~s$^{-1}$)$|$$\lesssim$100 appears located closer to the central regions and does not show a clear velocity gradient. For the other molecules the velocity gradient also becomes unclear for velocity offsets $|$v$_{\rm offset}$$|$$>$80~km~s$^{-1}$. This is probably due to some the molecular gas moving in a more complicated way than the observed velocity gradient. In order to compare the ALMA observations with the spatio-kinematical model of a radially-expanding disk, we used the software SHAPE \citep{Steffen2011} to visualize the resulting velocity field. The parameters of the model were obtained as follows. Assuming that the molecular gas has expanded with constant velocity since it was ejected, the inclination of the polar axis of the disk with respect to the line-of-sight, $i$\footnote{$i$$=$0$^{\circ}$ and $i$$=$90$^{\circ}$ correspond to the disk seen face-on and edge-on, respectively},  can be estimated from the following expression:
\begin{equation*}
\sin{i}=0.843\left[\frac{{\rm v}_{\rm edge}}{80\,{\rm km \, s}^{-1}}\right]\left[\frac{\tau}{100\,{\rm yr}}\right]\left[\frac{D}{4\,{\rm kpc}}\right]^{-1}\left[\frac{\theta_{\rm maj}}{1^{\prime\prime}}\right]^{-1}, \end{equation*}
\noindent where v$_{\rm edge}$ is the maximum observed value of the velocity offset at the edge of the disk, $\tau$ is the time since the ejection of the molecular gas, which is assumed to have occurred at the same time as the VLTP, $D$ is the distance to V\,605~Aql, and $\theta_{\rm maj}$ is the angular size of the major axis of the disk. The value of v$_{\rm edge}$ is obtained from the velocity offset at the edge of the disk in the first moment image, namely, v$_{\rm edge}$$=$80~km~s$^{-1}$. The angular size of the major axis of the disk is obtained from an elliptical fit to the CO($J$$=$3$\shortrightarrow$2) emission, which gives $\theta_{\rm maj}$$\approx$1$^{\prime\prime}$, with a P.A.$=$$-$37$^{\circ}$. This orientation is in excellent agreement with the direction of a dark band in the ejecta of V\,605~Aql seen at optical wavelengths \citep[see e.g.][]{Hinkle2008,Clayton2013}. Thus, for a distance to the source of 4 kpc, the derived inclination of the polar axis of the disk is $i$$\approx$60$^{\circ}$. Consequently, the de-projected expansion velocity of the disk is v$_{\rm exp}$$\approx$90~km~s$^{-1}$. The right panel of Figure~\ref{fig:moment1_model} shows a spatio-kinematical model created with SHAPE, which includes a radially-expanding disk inclined with respect to the line-of-sight, $i$$=$60$^{\circ}$, and with an expansion velocity v$_{\rm exp}$$=$90~km~s$^{-1}$. It is worth noting that the SHAPE modeling does not consider radiative transfer parameters and the output just shows the spatio-kinematical characteristics of the model, which are qualitatively in good agreement with the observations.  

In addition to the gradient discussed above, we found that the CO($J$$=$3$\shortrightarrow$2) emission with velocity offsets $-$120$\lesssim$v$_{\rm offset}$(km~s$^{-1}$)$\lesssim$$-$100 and +100$\lesssim$v$_{\rm offset}$(km~s$^{-1}$)$\lesssim$+140 exhibits an inverted velocity gradient, i.e. the blue- and red-shifted emission is located toward NE and SW from the center of the system, respectively (see left panel of Fig.~\ref{fig:moment1_model}). Since these components are the ones with the highest velocity offset, we refer to them as high velocity component (HVC). The HVC, arising from a bipolar outflow, is not clearly identified or isolated as such in the emission of the other molecules. Although the HCN line also shows broad emission wings, the lack of a clear velocity gradient suggests gas moving in a more complex way than a bipolar outflow.
The de-projected expansion velocity of the HVC, considering the inclination $i$$=$60$^{\circ}$ derived above, is $\approx$280~km~s$^{-1}$. The velocity gradient of the HVC is the one expected for a bipolar outflow oriented perpendicularly to the expanding disk. This is shown in the SHAPE model of Figure~\ref{fig:moment1_model} as two lobes along the polar axis of the disk. It should be noted that the spatio-kinematical components revealed by our ALMA observations are in complete agreement with the spatial configuration derived from optical observations of the bipolar lobes of V\,605~Aql, which show that emission from the SW-lobe exhibits a higher degree of absorption and is red-shifted, indicating that it lies behind an absorbing toroidal component.

A remarkable result from the ALMA observations is that, even though the HVC has the highest velocity offset, it lies just within $\approx$0$\rlap{.}^{\prime\prime}$2 ($\lesssim$1000~AU) from the center of the disk. Considering that the gas is moving at $\sim$280~km~s$^{-1}$, this implies a kinematical age for the bipolar outflow of $\lesssim$20 yr. This result would suggest that molecular material is currently being ejected from V\,605~Aql, and that we are witnessing the collimation of mass-loss by a born-again star even after 100~yr of experiencing a VLTP. An alternative explanation for the bipolar outflow is that it consists of molecular material that is being dragged by a stellar wind from the inner regions of the disk. 

In order to derive the energetics of the ejected material we corrected for the inclination and assuming optically thin emission for the CO($J$$=$3$\shortrightarrow$2) line and an excitation temperature $T_{\rm ex}$$=$75~K, we estimate that the scalar moment carried by the CO gas is 8$\pm$0.1$\times$10$^{36}$~gm~cm~s$^{-1}$ and the kinetic energy is 5$\pm$0.1$\times$10$^{43}$~erg. For comparison, we note that the kinetic energy of the hydrogen-deficient ejecta in other born-again PNe such as A\,30, A\,78 and HuBi\,1 is $\sim$10$^{44}$--10$^{45}$~erg.

\subsection{The $^{12}$C/$^{13}$C abundance ratio in V\,605~Aql} \label{sec:isotope_ratio}

The $^{12}$C/$^{13}$C abundance ratio can be estimated from the H$^{12}$CN ($J$$=$4$\shortrightarrow$3) and H$^{13}$CN ($J$$=$4$\shortrightarrow$3) lines, assuming that the emission is optically thin for both lines. Under this assumption, we obtain a H$^{12}$CN/H$^{13}$CN abundance ratio of 5.6$\pm$0.6, which is essentially independent of the excitation temperature. This value is in agreement with the lower limit, $^{12}$C/$^{13}$C$>$4, derived by \citet{Tafoya2017} from APEX observations.    

The H$^{12}$CN/H$^{13}$CN abundance ratio from our ALMA observations is consistent with the $^{12}$C/$^{13}$C abundance ratio value of $\approx$5.3 obtained from VLTP calculations presented in \citet{MB2006}, which suggest that the hydrogen-deficient ejection of material in V\,605~Aql was produced by the evolution of a single star, instead of an interacting binary system in a nova-like channel as proposed by \citet[][]{Lau2011}. However, if the HVC is indeed due to a collimated outflow, the presence of a binary companion could explain the launching of such an outflow. 

A possible scenario that could explain our ALMA observations is that when V\,605~Aql underwent the VLTP, which caused the expansion of its external layers, it experienced a common envelope event with a companion, such as the ones discussed by \citet[][]{Ivanova2013}. This would explain the formation of the toroidal component of the hydrogen-deficient material, i.e. the expanding disk, and the presence of the on-going bipolar ejection, i.e. the HVC. A similar scenario has been hinted by \citet{Toala2021} for the case of A\,30 in which a binary companion has been suggested to exist \citep[][]{Jacoby2020}.

\section{Final Remarks} \label{sec:final_remarks}
The high-angular ALMA observations have allowed us to image for the first time the spatial distribution and velocity field of the molecular emission in a born-again star. The clumpy ring-like morphology of the emission resembles the elliptical distribution of hydrogen-deficient clumps seen in the more evolved born-again PNe A30 and A78 \citep[e.g.][]{Toala2021}. On the other hand, the presence of high-density tracer molecules, such as HCN and HCO$^{+}$, suggest that the expanding molecular disk seen in V\,605~Aql contains dense neutral regions. Thus, it is expected that, after ionization and erosion by the energetic radiation and fast winds from the central star, the molecular disk will turn into a disrupted ring similar to the ones seen in later stages of the born-again star evolution. In addition, given the rapid evolution of born-again stars like V\,605~Aql, regular observations like the ones presented in this work will allow to witness the evolution and fate of the dust and molecular material produced in the VLTP. Finally, in order to better constrain the physical parameters of the dust and gas within the disk, observations of other molecular transitions and at other frequency bands are necessary.


\begin{acknowledgments}
This Letter makes use of the following ALMA data: ADS/ JAO.ALMA\#2019.1.01408.S. ALMA is a partnership of ESO (representing its member states), NSF (USA) and NINS (Japan), together with NRC (Canada), MOST and ASIAA (Taiwan), and KASI (Republic of Korea), in cooperation with the Republic of Chile. The Joint ALMA Observatory is operated by ESO, AUI/NRAO and NAOJ. JAT thanks funding by Fundación Marcos Moshinsky (M\'{e}xico) and the DGAPA UNAM projects IA100720 and IA101622. JBRG thanks Consejo Nacional de Ciencias y Tecnolog\'{i}a (CONACYT) Mexico for a research student grant. MAG acknowledges support of the Spanish Ministerio de Ciencia, Innovaci\'{o}n y Universidades (MCIU) grant PGC2018-102184-B-I00 and from the State Agency for Research of the Spanish MCIU through the ``Center of Excellence Severo Ochoa'' award to the Instituto de Astrof\'{i}sica de Andaluc\'{i}a (SEV-2017-0709).  LAZ acknowledges financial support from CONACyT-280775 and UNAM-PAPIIT IN110618 grants, M\'{e}xico. SPTM and acknowledges to Chalmers Gender Initiative for Excellence (Genie). This work has made extensive use of NASA's Astrophysics Data System. The authors thank Theo Khouri for helpful discussions. The authors also thank the anonymous referee for constructive comments and suggestions that helped to improve the manuscript.
\end{acknowledgments}



\software{CASA {\citep{McMullin2007}}, SHAPE \citep{Steffen2011}}
\facilities{ALMA, HST}



\FloatBarrier

\end{document}